\documentstyle[12pt]{article}

\newcommand{\be}{\begin{equation}}  
\newcommand{\ee}{\end{equation}}  
\newcommand{\beqa}{\begin{eqnarray}}  
\newcommand{\eeqa}{\end{eqnarray}}  
\newcommand{\beqar}{\begin{eqnarray*}}  
\newcommand{\eeqar}{\end{eqnarray*}}

\newcommand\gaml{\gamma_{ab}}

\newcommand\Tl{T_{ab}}  
  
\newcommand\Kl{K_{ab}}
\newcommand\dR{\dot{R}}

\begin{document}  
  
%\baselineskip 0.65cm  
%\title  
\rightline{EFI-99-43}  
\rightline{hep-th/9910149}  
\vskip 1cm  
\centerline{\Large \bf Dynamics of Anti-de Sitter Domain Walls}  
\vskip 1cm  
  
\renewcommand{\thefootnote}{\fnsymbol{footnote}}  
\centerline{{\bf Per Kraus${}$\footnote{pkraus@theory.uchicago.edu}}}   
\vskip .5cm  
 \centerline{ \it Enrico Fermi Institute,}  
\centerline{\it University of Chicago,}  
\centerline{\it Chicago, IL 60637, USA}  
  
\setcounter{footnote}{0}  
\renewcommand{\thefootnote}{\arabic{footnote}}  
  
\begin{abstract} 

We study solutions corresponding to moving domain walls in  the 
Randall-Sundrum universe.  The bulk geometry is given by patching 
together black hole solutions in AdS$_5$, and the motion of the wall is 
determined from the junction equations.  Observers on the wall interpret
the motion as cosmological expansion or contraction.  We describe the 
possible wall trajectories, and examine the consequences for localized
gravity on the wall.  
  
\end{abstract}

\setcounter{footnote}{0}

\section{Introduction}   
\baselineskip 0.60cm 
Randall and Sundrum \cite{RS} have recently discovered a novel way of embedding
four dimensional physics within a higher dimensional world.  In their
setup a four dimensional domain wall sits at a point in an infinite but 
highly curved fifth dimension, and a single normalizable zero mode of 
the gravitational field gives rise to Newtonian gravity at large distances
on the wall.  The geometry as a whole is that of two regions of AdS$_5$
joined by the domain wall. 

The presence of four dimensional Poincar\'{e} invariance requires a precise
value for the domain wall tension.  However, it is also of interest to
consider  non-Poincar\'{e} invariant solutions, both to better 
understand the mechanism of localized gravity and for possible 
cosmological applications.  When the tension is not fine-tuned or
there is additional matter on the wall, time dependent solutions typically
result.  Loosely speaking, 
such time dependence can come in two forms.  As studied in
\cite{Kaloper:1999sm,Nihei:1999mt,Csaki:1999jh,Cline:1999ts,Kim:1999ja,DeWolfe:1999cp,Kanti:1999sz,Cline:1999tq}, 
one can find cosmological solutions in which the bulk geometry 
is time dependent.  Here we study the alternative case  in which the 
bulk remains static but the domain wall acquires a velocity.  Observers
on the wall will interpret the motion of the wall through the static
background as cosmological expansion or  contraction.  More generally,
one can combine the two forms of time dependence to find a large class 
of solutions, but that will not be considered here.   Of course, given
any moving domain wall one can always transform coordinates to put it
at rest, hence there
is  some overlap with the solutions found in
\cite{Kaloper:1999sm,Nihei:1999mt,Csaki:1999jh,Cline:1999ts,Kim:1999ja,DeWolfe:1999cp,Kanti:1999sz,Cline:1999tq}
   and those described here.  

The equations of motion of the wall are found by a straightforward
application of the thin wall formalism in general relativity~\cite{israel}, 
in which Einstein's
equations are rewritten as junction conditions relating the discontinuity
in the wall's extrinsic curvature to its energy-momentum tensor.  We will
consider the  general solutions which have the symmetries of the
standard Robertson-Walker geometries.  Depending on the choice of 
parameters, we find   bounded and unbounded wall trajectories, with both
exponential and power law expansion in the latter case.  

Our solutions are  generally not described by regions of AdS$_5$ in the
bulk, but rather by black hole solutions which reduce to AdS$_5$ as
a non-extremality parameter is taken to zero.  This has interesting 
implications for solutions in which the domain wall is slowly moving, so
that one might expect to recover four dimensional gravity as in 
\cite{RS}.  If taken to be too large, the event horizon of the black 
hole will render the zero mode nonnormalizable, and so destroy the
effective four dimensional behavior.  Small horizons result in zero 
modes, but the bulk solution has reduced symmetry compared to AdS$_5$.
The lack of Lorentz invariance in the bulk manifests itself as a shift
in the propagation speed of gravitational fluctuations on the wall,  
although the shift rapidly becomes negligible as the wall universe expands.

Eventually, one would like a microscopic description of the domain wall,
either as a smooth solution of supergravity (along the lines of 
\cite{Cvetic:1996vr,Behrndt:1999kz,Skenderis:1999mm}) or as a fundamental brane of string theory.  
The present analysis is applicable to the former case in the thin wall limit.  
As an attempt to realize the latter we conclude by describing
 a configuration involving
a spherical distribution of D3-branes, which ends up being unsuccessful
due to the low value of the D3-brane tension.

Dynamical domain walls have also been studied recently 
in~\cite{Chamblin:1999ya,Chamblin:1999ea}.

\section{Junction equations}   
\baselineskip 0.60cm 

The five dimensional bulk gravitational action is\footnote{Five
dimensional indices are denoted by $\mu=0\ldots 4$, four dimensional
indices on the domain wall by $a=0 \dots 3$, and spatial
indices on the wall by $i=1 \ldots 3$.}
\be
  S= \frac{1}{16\pi G}\int_M d^{5}x \sqrt{- g}
  \left( R + \frac{12}{\ell^2}\right)
  +\frac{1}{8\pi G}\int_{\partial M} d^4 x \sqrt{-\gamma} ~K~.
\label{action}
\ee
We have allowed for a boundary $\partial M$ with induced metric
$\gamma_{ab}$, shortly to be identified with the domain wall.  $K$
is the trace of the extrinsic curvature of the boundary
\be
K^{\mu\nu} = \nabla^\mu n^\nu,
\ee
where $n^\mu$ is the  unit normal vector on $\partial M$. 
We are interested in patching two such regions together across a domain
wall with four dimensional energy-momentum tensor $T_{ab}$.  It is
convenient to work in Gaussian normal coordinates near the domain wall,
\be
ds^2 = \gaml dx^a dx^b + d\eta^2,
\label{gaussian}
\ee
with the wall at $\eta=0$.  Then Einstein's equations imply the
junction conditions~\cite{israel}:
\be
\Delta \Kl = -8\pi G \left(\Tl - {1 \over 3}T_c^c \gaml \right),
\ee
where
\be
\Delta \Kl =\Kl^+ - \Kl^- = \Kl(\eta= \epsilon) - \Kl(\eta = -\epsilon).
\ee
The relative minus sign arises because we have chosen the convention that
$n^\mu$ points towards the region of increasing $\eta$.  In the
coordinates (\ref{gaussian}) the extrinsic curvature is
\be
\Kl = {1 \over 2} \partial_{\eta} \gaml.
\label{gauscurv}
\ee
We refer to an ``extremal'' wall as one with energy-momentum tensor
$\Tl = -\sigma \gaml$, in which case the junction conditions become
\be
\Delta \Kl = -{8 \pi G \sigma \over 3} \gaml.
\ee
%
%\subsection{Bulk Spacetimes}
%

We will consider bulk solutions that have the symmetries of flat, open,
and closed Robertson-Walker universes.   The  unique solutions of Einstein's
equations with the assumed properties are
\be
ds^2 = -(k+{r^2 \over \ell^2} - {\mu \over r^2} ) dt^2
+r^2 d\Sigma_k^2 + {dr^2 \over (k+{r^2 \over \ell^2} - {\mu \over r^2} )}.
\label{metric}
\ee
$k$ takes the values $0$, $-1$, $+1$, corresponding to flat, open, or
closed geometries, and $d\Sigma_k^2$ is the corresponding metric on the
unit three dimensional plane, hyperboloid, or sphere.  The $k=-1$ solution
is perhaps unfamiliar, but has been studied 
in~\cite{Birmingham:1998nr,Emparan:1999pm,Emparan:1999gf}. When 
$\mu =0$ the solutions are simply AdS$_5$ written in various coordinates,
whereas $\mu \neq 0$ gives black hole solutions with horizons at $r=r_h$,
\be
r_h^2 = {\ell^2 \over 2}(-k + \sqrt{k^2 + 4 \mu/\ell^2}).
\ee
$k=0,1$ requires $\mu \geq0$, and $k=-1$ requires $\mu \geq -\ell^2/4$.
Note that the five dimensional version of Birkhoff's theorem requires
that the solutions (\ref{metric}) be static.  

The domain wall separating two such spacetimes (with the same $k$) is 
taken to be situated at $r=R(t)$, where $R(t)$ will be determined by
solving the junction equations.
%We will patch two such spacetimes (with the same $k$) together across a
%domain wall located at $r=R(t)$, where $R(t)$ will be determined by
%solving the junction equations.  
As in the Randall-Sundrum geometry, $r$
is taken to decrease as one moves away from the wall in {\em either} 
direction.  
One way to determine the junction equations is to transform to 
Gaussian normal coordinates and use the formula (\ref{gauscurv}). 
However, it is simpler to rewrite the equations in a coordinate 
independent form.  Let $u^\mu$ be the velocity vector of the wall,
$u^\mu u_\mu =-1$.  Then the unit normal satisifies $n^\mu u_\mu =0$,
and we can rewrite (\ref{gauscurv}) as
\be
\Kl = {1 \over 2} n^\mu \partial_\mu \gaml.
\label{junction}
\ee

Let us apply this to the metric
\be
ds^2 = -f_k(r) dt^2 + r^2 d\Sigma_k^2 + f_k^{-1}(r) dr^2.
\ee
We have $u^t = (f_k + \dR^2)^{1/2}f_k^{-1}$, $u^r = \dR$, where 
$\dR$ is the derivative of $R$ with respect to proper time $\tau$.  
Then $n^t = -f_k^{-1} \dR$, $n^r = -(f_k + \dR^2)^{1/2}$.  The minus sign
arises because the coordinate $r$ is decreasing in the direction $n^\mu$. 
There are two nontrivial junction equations corresponding to the time and
space components of  (\ref{junction}).  The spatial components of 
the extrinsic curvature are 
\be
K_{ij}^+ = - {(f_k^+ + \dR^2)^{1/2} \over R} \gamma_{ij}, \quad\quad
K_{ij}^- =  {(f_k^- + \dR^2)^{1/2} \over R} \gamma_{ij},
\ee
wher $\pm$ denotes the two sides of the wall.
% and the relative minus
%sign arises from the convention that $n^\mu$ points towards the $+$ region. 
The junction equation is then
\be
\left[(f_k^+ + \dR^2)^{1/2}+(f_k^- + \dR^2)^{1/2}  \right]\gamma_{ij}
= 8\pi G R(T_{ij} - {1 \over 3}T^a_a \gamma_{ij}),
\ee
or in the extremal case ($\Tl = -\sigma \gaml$):
\be
(f_k^+ + \dR^2)^{1/2}+(f_k^- + \dR^2)^{1/2} = {8 \pi G \sigma \over 3 } R.
\label{exteq}
\ee
It turns out that the junction equation for $K_{tt}$ just gives the proper 
time derivative of the equations above, so we need not consider it further. 

The junction equation determines $R(\tau)$, and so also the induced metric
on the domain wall:
\be
ds_{wall}^2 = -d\tau^2 + R^2(\tau) d\Sigma_k^2.
\ee

For reference, recall that in the standard FRW cosmology with energy density
$\rho$,   the scale factor $R$ obeys the equation
\be
{1 \over 2} \dR^2 - {4 \pi G_N \rho  \over 3}R^2 = -k/2.
\label{FRW}
\ee

\subsection{Motion of extremal wall}

We can rewrite (\ref{exteq}) as the equation for a particle in a potential
\be
{1 \over 2} \dR^2 +V(R) = -{k \over 2},
\label{eqmo}
\ee
with 
\be
V(R) = {1 \over 2} \left( 1- \left({\sigma \over \sigma_c}\right)^2 \right)
{R^2 \over \ell^2} - {1 \over 4} {(\mu_+ + \mu_-) \over R^2}
- {1 \over 32} \left({\sigma \over \sigma_c}\right)^{-2} { \ell^2
(\mu_+ - \mu_-)^2 \over R^6},
\ee
where we have defined $\sigma_c \equiv 3/(4\pi G \ell)$.
The Randall-Sundrum configuration results from taking $k=\mu_+ =\mu_- =0$,
and tuning the wall tension to be $\sigma = \sigma_c$.
We now explore some of the possibilities that arise when we relax these
conditions. 

First consider the case $\mu_+ =\mu_- =0$, in which the bulk geometries
are regions of AdS$_5$.   There are nine cases corresponding to the
various values of $\sigma$ and $k$:
\vskip 0.3cm
\noindent
\underline{ $\sigma = \sigma_c$, $k=0$:}  ~This gives the Randall-Sundrum
configuration.
\vskip 0.3cm
\noindent
\underline{ $\sigma = \sigma_c$, $k=-1$:}~ $R(\tau) = |\tau|$.
Wall passes  between $R=0$ and $R = \infty$, and crosses the 
coordinate horizon $r=\ell$ in finite proper time.  
\vskip 0.3cm
\noindent
\underline{ $\sigma = \sigma_c$, $k=+1$:} ~ No solution.
\vskip 0.3cm
\noindent
\underline{ $\sigma > \sigma_c$, $k=0$:}~
$R = R_0 e^{\pm H\tau}$. 
 \vskip 0.3cm
\noindent
\underline{ $\sigma > \sigma_c$, $k=-1$:}~$R=H^{-1}
\sinh{H \tau}$. 
\vskip 0.3cm
\noindent
\underline{ $\sigma > \sigma_c$, $k=+1$:}~$R=H^{-1}
\cosh{H \tau}$.
\vskip 0.3cm
\noindent
\underline{ $\sigma < \sigma_c$, $k=0$:}~ No solution.
\vskip 0.3cm
\noindent
\underline{ $\sigma < \sigma_c$, $k=-1$:}~ $R=H^{-1}
\cos{H \tau}$. 
\vskip 0.3cm
\noindent
\underline{ $\sigma < \sigma_c$, $k=+1$:}~ No solution.

\vskip 0.4cm

Here $H= {1 \over \ell}\sqrt{|(\sigma/\sigma_c)^2-1|}$.   In the three
$\sigma > \sigma_c$ cases the wall metric is de-Sitter space, while in 
the $\sigma < \sigma_c$ case it is anti-de Sitter space.  These 
solutions have appeared in different coordinates in the work of
\cite{Kaloper:1999sm,Nihei:1999mt,Csaki:1999jh,Cline:1999ts,Kim:1999ja,DeWolfe:1999cp,Kanti:1999sz,Cline:1999tq}.

Now let us turn to the case where $\mu_+ , \mu_- \neq 0$.  The cases
$\sigma > \sigma_c$, $\sigma < \sigma_c$ are qualitatively similar
to those described above, either inflationary behavior for large $R$ or
bounded motion.  Note, though, that the possibility of $\mu_+ + \mu_- < 0$ 
for $k=-1$ allows $V(R)$ to have nontrivial local maxima. 
The detailed forms of the trajectories can be found by integrating
the equation of motion~(\ref{eqmo}). 
  We now
consider the three $\sigma =\sigma_c$ cases:
\vskip 0.3cm
\noindent
\underline{ $\sigma = \sigma_c$, $k=0$:}  ~Unbounded motion passing
between $R=0$ and
$R(\tau) \approx 2^{1/4}(\mu_+ + \mu_-)^{1/4} |\tau|^{1/2}$.   
For large $R$ the
wall metric is that of a spatially flat radiation dominated cosmology.
\vskip 0.3cm
\noindent
\underline{ $\sigma = \sigma_c$, $k=-1$:}  ~Unbounded motion passing between
$R=0$ and $R(\tau) \approx |\tau|$. 
\vskip 0.3cm
\noindent
\underline{ $\sigma = \sigma_c$, $k=+1$:}  ~Wall expands from $R=0$ to 
maximum size, $V(R_{max})=-1/2$, and recollapses.
\vskip 0.4cm

\subsection{Four dimensional description}

It is interesting to consider the case 
$\sigma=\sigma_c$, $k=0$, $\mu_+ , \mu_- \neq 0$ in more 
detail.  At late times, $\tau \gg (\mu_+ + \mu_-)^{1/2}$, the wall
universe is slowly expanding and it becomes meaningful to ask whether,
as in \cite{RS},
conventional gravity in four approximately flat spacetime dimensions is
recovered for distances large compared to $\ell$ but small compared to 
$(\dR/R)^{-1}$.   The latter condition means that we can take
$R$ to be constant over the time scale of interest.  For simplicity, take
$\mu_+ = \mu_- = \mu$, so that at late times the bulk geometry is that
of the Randall-Sundrum configuration except that the bulk spacetime is
the black hole geometry (\ref{metric}).  There are two important new
features: the infinite throat as $r \rightarrow 0$ has been replaced by
an event horizon at $r=r_h = \ell^{1/2} \mu^{1/4}$, and four dimensional
Lorentz invariance has been broken.  To study the implications in a 
simplified setting we will replace the gravitational fluctuations with
those of a massless bulk scalar field.  When $r_h$ is set to zero, 
the scalar field has an $r$ independent normalizable zero mode, 
$\phi = \phi(t,\vec{x})$, which appears as a massless four dimensional 
scalar field on the domain wall.  We can study the fate of the zero mode
by examining the wave equation near the horizon using the coordinate
$r_* = (\ell^2 / 4 r_h) \ln( (r-r_h)/\ell)$.  Writing
$\phi = e^{-i\omega t + i \vec{k} \cdot \vec{x}} \psi(r_*)$ the wave
equation becomes
\be
\left(\partial_{r_*}^2 + \omega^2 - {4 \over \ell r_h} e^{4 r_h r_* / \ell^2} 
\vec{k}^2 \right) \psi(r_*)=0. 
\ee
We see that there is no normalizable mode for $\omega \neq 0$, which seems to 
imply the lack of a massless field on the wall. On the other hand, we 
know that such a mode exists for $r_h=0$ and we expect the limit
$r_h \rightarrow 0$ to be smooth.  The resolution is that for small $r_h$ the
geometry near the horizon is not reliable, and so we should impose a cutoff on
the range of $r$.  To implement this we work out the action of the candidate
zero mode $\phi = \phi(t,\vec{x})$:
\be
 S = - {1 \over 2} \int \! d^5x \, \sqrt{-g} g^{\mu\nu}\partial_\mu
\phi \partial_\nu \phi 
=-  \int \! d^4x \, \int_{r_c}^R \! dr \, r^3
\left[ - {(\partial_t \phi)^2 \over ({r^2 \over \ell^2} - {\mu \over r^2})}
+ {(\vec{\nabla}\phi)^2 \over r^2} \right].
\ee 
Now, the trouble arises from attempting to take $r_c \rightarrow r_h$; 
instead, for small $r_h$ we impose the cutoff $r_c = \ell$ corresponding
to the region where trans-Planckian curvatures begin to set in (we have in
mind that $\ell \approx \ell_{{\rm Pl}}$).   Then evaluating the integrals and
expanding in $r_h$ we find, assuming $R \gg \ell$:
\be
S= - {R^2 \over 2} \int \! d^4x \, 
\left[-\ell^2 \left(1+{r_h^4 \over \ell^2 R^2}\right)
\left(\partial_t \phi \right)^2 +(\vec{\nabla}\phi)^2 \right].
\ee
Rescaling $\phi$ and expressing the result in terms of the domain
wall metric $\gaml$ we obtain:
\be
S=-{1 \over 2} \int \! d^4x \, \sqrt{-\gamma} \left[
\left(1+{r_h^4 \over \ell^2 R^2}\right)
\gamma^{tt}\left(\partial_t \phi \right)^2 + \gamma^{ij}
\partial_i \phi \partial_j \phi \right],
\ee
which is the standard form except that the speed of light has been shifted
to
\be
c_{eff} \approx 1 - {1 \over 2} {r_h^4 \over \ell^2 R^2}.
\ee
We stress that this formula holds only when the correction term is small,
and that the precise correction is not meaningful since it is sensitive
to the position of the cutoff.  

If the standard model fields live on the domain wall then their
behavior is Lorentz invariant with the standard speed of light $c=1$,
whereas gravitational interactions (assuming that the scalar field
results can be extrapolated to gravity) propagate at a slightly shifted
speed due to the loss of Lorentz invariance in the bulk.  In addition,
we expect  the tensor and momentum structure of the gravitational
interactions to  suffer small Lorentz violating corrections.  As the universe
expands, $R$ becomes large and these effects rapidly become negligible. 

Finally, in the case where $r_h$ is large compared to $\ell$, we expect the 
four dimensional description to be invalid due to the lack  of a normalizable
zero mode.   

\subsection{Matter on the wall}

Now consider the case in
 which expansion results from matter on the domain wall.
We take for the  energy-momentum tensor
\be
\Tl = -\sigma_c \gaml + \rho u_a u_b + p (\gaml + u_a u_b),
\ee
corresponding to matter with energy density $\rho$ and pressure $p$, in
addition to the critical background wall tension $\sigma_c$.    
Energy conservation requires 
\be
{d \over d\tau}(\rho R^3) = -p {d \over d\tau} R^3.
\ee
 The junction equation is found to be
\be
(f_k^+ + \dR^2)^{1/2}+(f_k^- + \dR^2)^{1/2} 
= {8 \pi G (\sigma_c +\rho) \over 3 } R.
\ee
For $\rho \ll \sigma_c$ the corresponding potential is
\be
V(R) = -{\rho R^2 \over \ell^2 \sigma_c }
 - {1 \over 4} {(\mu_+ + \mu_-) \over R^2}
- {1 \over 32} \left(1-{2 \rho \over \sigma_c }\right) { \ell^2
(\mu_+ - \mu_-)^2 \over R^6}.
\ee
With $\mu_+ = \mu_- =0$ the resulting potential is the same as that 
which arises in standard FRW cosmology (\ref{FRW}), hence one recovers the 
conventional behavior.  In particular, a $k=0$ radiation dominated 
universe with $p=\rho/3$, $\rho = \rho_0/R^4$, leads to 
$R(\tau) = (4\rho_0 / \ell^2 \sigma_c)^{1/4} |\tau|^{1/2}$.  A 
$k=0$ matter dominated 
universe with $p=0$, $\rho = \rho_0/R^3$ leads to 
$R(\tau) = (9 \rho_0 / 2\ell^2 \sigma_c)^{1/3} |\tau|^{2/3}$.

When $\mu_+, \mu_- \neq 0$, the late time behavior of a 
radiation dominated $k=0$ universe
again has $R(\tau) \sim const \cdot |\tau|^{1/2}$, but with a 
shifted effective energy density: $\rho_0 \rightarrow \rho_0 + \ell^2 \sigma_c
(\mu_+ + \mu_- )/4$.   On the other hand, the late time behavior of
a matter dominated $k=0$ universe is unchanged.  

\section{Attempt at a string theory realization}

It is desirable to have an embedding of the Randall-Sundrum geometry into
string theory.  Here we briefly describe a largely unsuccessful attempt
based on a spherical shell of 
D3-branes~\footnote{See~\cite{Kraus:1998hv,Kehagias:1999ju,Brandhuber:1999hb,Bakas:1999ax,Grojean:1999uu} for related discussions.}.  
The construction fails because
the tension of a D3-brane turns out to be too small by a factor of
$3/2$.  

As is well known, the near horizon geometry of a collection of D3-branes
is AdS$_5 \times S^5$,
\be
ds^2 = -{r^2 \over \ell^2} dt^2 + r^2(d \vec{x})^2 + {\ell^2 \over r^2} dr^2
+ \ell^2 d\Omega_5^2.
\ee
$\ell$ is related to the number of D3-branes by 
$\ell^4 = 4\pi g N (\alpha')^2$.  In addition, there are $N$ units of 
five-form flux present.  Now, we attempt to patch two such regions with
opposite five-form orientations together along a boundary of constant $r$.
To satisfy charge conservation we need the boundary to carry $2N$ units of 
charge.  We do this while preserving approximate $SO(6)$ symmetry by 
distributing $2N$ D3-branes over the $S^5$.  The branes are at coincident
$r$ positions, and their worldvolumes span $t, \vec{x}$.  To preserve
approximate $SO(6)$ symmetry we require that the inter-brane spacing
on $S^5$ be much smaller than the characteristic scale of the geometry $\ell$.
This requires $N \gg 1$.  

Now, the tension of $2N$ D3-branes is, in terms of the five dimensional 
Newton constant,
\be
\sigma = {1 \over 2 \pi G \ell} = {2 \over 3 } \sigma_c.
\ee
Hence the tension is too low to patch two such AdS$_5 \times S^5$ regions
together in this manner.  The only possibility is a time dependent 
$k=-1$ solution as discussed earlier.  Such a solution collapses to a 
singularity in a time scale of order $\ell$.

\vspace{0.2in}  
  
\paragraph{Acknowledgments: }
 Supported by NSF Grant  
No. PHY-9600697. I have benefitted from discussions with J. Harvey, 
F. Larsen, E. Martinec, and R. Sundrum.

%%Bibliography  

\end{document}